\documentstyle[editedvolume]{crckapb}  

\def\lapp{\ifmmode\stackrel{<}{_{\sim}}\else$\stackrel{<}{_{\sim}}$\fi}
\def\gapp{\ifmmode\stackrel{>}{_{\sim}}\else$\stackrel{>}{_{\sim}}$\fi}

\begin{opening}
\title{Neutron Star Birth Rates}
\author{D.R. Lorimer}
\institute{National Astronomy and Ionospheric Center\\
           Arecibo Observatory\\
	   HC3 Box 53995, Arecibo PR 00612, USA}
\end{opening}
\runningtitle{Neutron Star Birth Rates}

\begin{document}
\begin{figure}[hbt]
\setlength{\unitlength}{1in}
\begin{picture}(0,0)
\put(0,2.6){\includegraphics{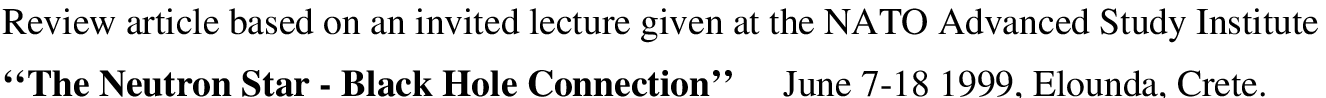}}
\end{picture}
\end{figure}

\begin{abstract}
A crucial test any proposed evolutionary scenario must pass is {\sl
can the birth rate of the sources we see be sustained by the proposed
progenitor population?} In this review, I investigate the methods used
to determine the birth rates of normal and millisecond radio pulsars
and summarise recent results for these two distinct neutron star populations.
\end{abstract}

\section{Introduction}
\label{sec:intro}

In principle, estimating the birth rate of a population of sources is
trivial: divide the total number of sources by the mean lifetime of
the population. In practice, however, for the neutron star population,
precise estimates of both the number and lifetime of the sources are
hard to obtain. The sample of Galactic neutron stars is heavily biased
by a number of observational selection effects which must be properly
accounted for in a birth rate analysis.

In this review I shall be mainly concerned with the birth rates of two
distinct sub-populations of the neutron star zoo: the normal pulsars
and the millisecond pulsars. \S \ref{sec:psrpop} defines what is meant
by these two classes. \S \ref{sec:evol} reviews the ``standard''
evolutionary scenarios which connect normal pulsars to supernova
remnants and millisecond pulsars to low-mass X-ray binaries. \S
\ref{sec:selfx} discusses the various selection effects known to
significantly bias these populations.  \S \ref{sec:corrfx} summarises
techniques to correct for these effects. This leads naturally to an
estimate of the birth rates (\S \ref{sec:rates}) which are discussed
in context with the proposed progenitor populations in \S \ref{sec:disc}. 
Some suggestions for future work are also made in this section.

\section{Normal and Millisecond Pulsars}
\label{sec:psrpop}

\begin{figure}[hbt]
\setlength{\unitlength}{1in}
\begin{picture}(0,3.4)
\put(0.3,4.1){\includegraphics{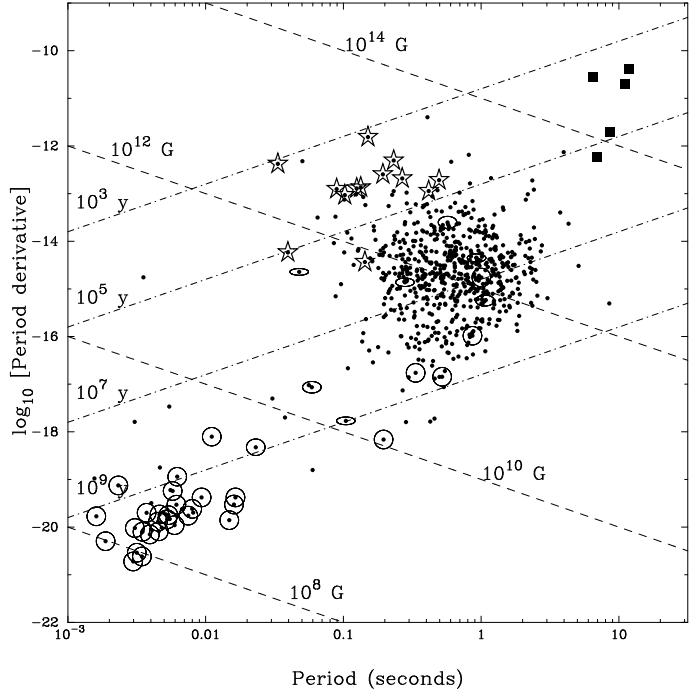}}
\end{picture}
\caption[]
{
The ubiquitous $P - \dot{P}$ diagram shown for a sample consisting of
radio pulsars (the black dots) and anomalous X-ray pulsars (the black
squares --- see review by Mereghetti in this volume). Pulsars known to
be members of binary systems are highlighted by a circle (for
low-eccentricity orbits) or an ellipse (for elliptical orbits). Pulsars 
thought to be associated with supernova remnants are highlighted by the 
starred symbols.
}
\label{fig:ppdot}
\end{figure}

As I alluded to in my other contribution to this volume, the period
$P$ and period derivative $\dot{P}$ for each pulsar can be obtained to
very high levels of precision through a series of timing measurements.
Perhaps the most oft-plotted figure from these data is the
``$P$---$\dot{P}$ diagram'' shown in Fig.~\ref{fig:ppdot}.

Amongst other things, the diagram demonstrates clearly the distinction
between the ``normal pulsars'' ($P \sim 0.5$ s and $\dot{P} \sim
10^{-15}$ s/s and populating the ``island'' of points) and the
``millisecond pulsars'' ($P \sim 3$ ms and $\dot{P} \sim 10^{-20}$ s/s
and occupying the lower left part of the diagram).

The differences in $P$ and $\dot{P}$ imply fundamentally different
typical ages and magnetic field strengths for the two populations. For
the ``canonical neutron star'' of mass 1.4 M$_{\odot}$, radius 10 km
and moment of inertia $10^{45}\,$g$\,$cm$^{2}$ (see for example
Shapiro \& Teukolsky 1983), it can be shown that the surface dipole
magnetic field strength in units of $10^{12}$ Gauss, $B_{12} \simeq
(P_s \, \dot P_{-15})^{1/2}$, where $P_s$ is the period in seconds and
$\dot P_{-15} = 10^{15} \dot P$.  Integrating this first-order
differential equation over time, assuming a constant magnetic field,
results in an expression for the age of the pulsar $t = \tau \left [ 1
- (P_0/{P})^2 \right ]$, where $\tau = P/2\dot{P}$ is the so-called
``characteristic age'' and $P_0$ is the period of the pulsar at $t=0$.
Under the assumption that the neutron star has slowed down
significantly since birth ($P_0 \ll P$), the characteristic age $\tau$
is a good approximation to the true age $t$.  Lines of constant $B$
and $\tau$ are drawn on Fig.~\ref{fig:ppdot}.  From these, we infer
typical magnetic fields and ages of $10^{12}$ G and $10^{7}$ yr for
the normal pulsars and $10^{8}$ G and $10^{9}$ yr for the millisecond
pulsars respectively.

Whilst the consensus of evidence generally supports the assumptions in
the characteristic age estimates, viz: dipolar spin down from a small
initial spin period and a constant magnetic field (see
e.g.~Bhattacharya et al.~1992; Lorimer et al.~1993), we should caution
that, rather like dispersion-measure-derived distances, individual
characteristic ages should not be taken as precise values. It is
therefore of interest to look for independent age constraints. Strong
support of the characteristic age estimates for old pulsars comes from
the observations of asymmetric drift seen in the proper motions of
normal pulsars with large characteristic ages (Hansen \& Phinney 1997), and
overwhelmingly in the millisecond pulsars (Toscano et al.~1999). 
See Ramachandran's review in this volume for further discussion.

A very important additional difference between normal and millisecond
pulsars is binarity.  Orbital companions are much more commonly
observed around millisecond pulsars ($\sim 80$\% of the observed
sample) than the normal pulsars ($\lapp 1$\%).  As discussed in my
other contribution to these proceedings, timing measurements constrain
the masses of orbiting companions which, when supplemented by
observations at other wavelengths, tell us a great deal about their
nature. The present sample of orbiting companions are either white
dwarfs, main sequence stars, or other neutron stars.  Binary pulsars
with low-mass companions ($\lapp 0.5$ M$_{\odot}$ --- predominantly
white dwarfs) usually have millisecond spin periods and essentially
circular orbits: $10^{-5} \lapp e \lapp 10^{-1}$. Measurements of the
``cooling ages'' of the white dwarfs (see e.g.~van Kerkwijk 1996)
provide further evidence that millisecond pulsars have typical ages of
a few Gyr. The binary pulsars with high-mass companions ($\gapp 1$
M$_{\odot}$ --- neutron stars or main sequence stars) have larger spin
periods ($\gapp 30$ ms) and are in much more eccentric orbits: $0.2
\lapp e \lapp 0.9$.

\section{Evolutionary Scenarios}
\label{sec:evol}

We now briefly review the various end-products that are implied by
standard models for the formation and evolution of single and binary
radio pulsars which basically assume that every neutron star in the
disk of our Galaxy was formed during the core-collapse phase of a
supernova explosion of a massive star (see also van den Heuvel's
contribution to this volume).

\begin{figure}[hbt]
\setlength{\unitlength}{1in}
\begin{picture}(0,0.75)
\put(0.05,0){\includegraphics{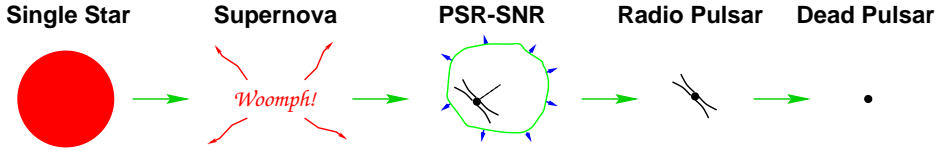}}
\end{picture}
\caption[]{
Cartoon showing the evolutionary sequence of a single neutron star.
}
\label{fig:sevol}
\end{figure}

The simplest scenario is shown in Fig.~\ref{fig:sevol} and begins in
the final moments of the life of a single massive star that is about
to explode as a supernova.
The neutron star formed during the core collapse will receive an
impulsive kick (see e.g. Spruit \& Phinney 1998) if the explosion is
not symmetric and, as a result, begin to move away from the centre of
the explosion. In the meantime, the outer layers of the star are
expanding into the surrounding space at velocities of up to 10,000 km
s$^{-1}$. The result is a pulsar--supernova remnant association
(PSR-SNR) which may be visible as a pair for up to $10^5$ yr after the
explosion. Eventually, the expanding shell becomes so diffuse that it
is no longer visible as a supernova remnant. The pulsar, on the other
hand, may produce radio emission for a further $10^7$ yr or more as it
gradually spins down to longer periods. Presently, the longest period
for a radio pulsar is 8.5 s (Young, Manchester \& Johnston 1999).  At
some point the rotational energy of the neutron star is insufficient
to induce pair production in its magnetosphere and the radio emission ceases.

\begin{figure}[hbt]
\setlength{\unitlength}{1in}
\begin{picture}(0,4.3)
\put(0.3,-0.05){\includegraphics{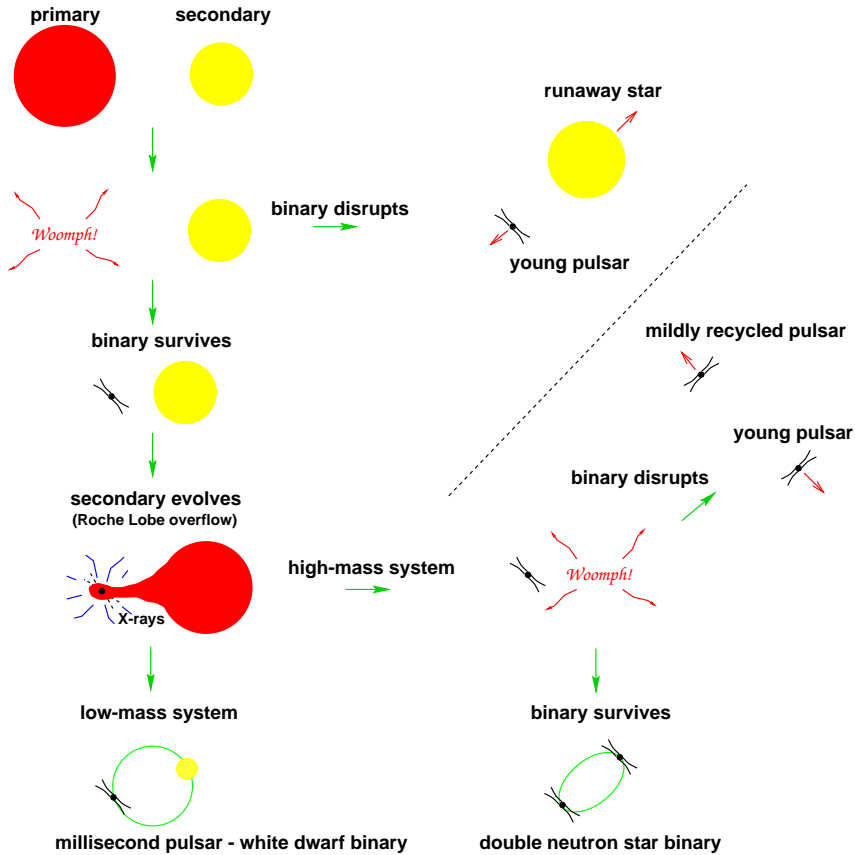}}
\end{picture}
\caption[]{
Cartoon showing various evolutionary scenarios involving binary pulsars.
}
\label{fig:bevol}
\end{figure}

The presently favoured model to explain the formation of the various
types of binary systems has been developed over the years by a number of
authors (Bisnovatyi-Kogan \& Komberg 1974; Flannery \& van den Heuvel
1975; Smarr \& Blandford 1976; Alpar et al.~1982). The model is
sketched in Fig.~\ref{fig:bevol} and can be qualitatively summarised
as follows: starting with a binary star system, the neutron star is
formed during the supernova explosion of the initially more massive
star (the primary) which has an inherently shorter main sequence
lifetime. From the virial theorem, it follows that the binary system
gets disrupted if more than half the total pre-supernova mass is
ejected from the system during the explosion.  In addition, the
fraction of surviving binaries is affected by the magnitude and
direction of the impulsive kick velocity the neutron star receives at
birth (Hills 1983; Bailes 1989).  Those binary systems that disrupt
produce a high-velocity isolated neutron star and an OB runaway star
(Blaauw 1961). Rather like the fortunate survivor of a crash, the
isolated pulsar has no recollection of the binary system it once
belonged to and behaves from the moment of release as a single pulsar
discussed above. The high binary disruption probability explains why
so few normal pulsars are observed with orbiting companions.

For those few binaries that remain bound, and where the companion star
is sufficiently massive to evolve into a giant and overflow its Roche
lobe, the old spun-down neutron star can gain a new lease of life as a
pulsar by accreting matter and therefore angular momentum from its
companion (Alpar et al.~1982). The term ``recycled pulsar'' is often
used to describe such objects. During this accretion phase, the X-rays
liberated by heating the material falling onto the neutron star mean
that such a system is expected to be visible as an X-ray binary system.

Two classes of X-ray binaries relevant to binary and millisecond
pulsars exist: neutron stars with high-mass or low-mass companions.
The high-mass companions are massive enough to explode themselves as a
supernova, producing a second neutron star. If the binary system is
lucky enough to survive this explosion, it ends up as a double neutron
star binary. The classic example is PSR B1913+16, a 59-ms radio pulsar
with a characteristic age of $\sim 10^8$ yr which orbits its companion
every 7.75 hr (Hulse \& Taylor 1975).  In this formation scenario, PSR
B1913+16 is an example of the older, first-born, neutron star that has
subsequently accreted matter from its companion. Presently, there are
no clear observable examples of the second-born neutron star in these
systems. This is probably reasonable when one realises that the
observable lifetimes of recycled pulsars are much larger than normal
pulsars.  As discussed by Kalogera in this volume, double neutron star
binary systems are very rare in the Galaxy --- another indication that
the majority of binary systems get disrupted when one of the
components explodes as a supernova.  Systems disrupted after the
supernova of the secondary form a midly-recycled isolated pulsar and a
young pulsar (formed during the explosion of the secondary).

By definition, the companions in the low-mass X-ray binaries evolve
and transfer matter onto the neutron star on a much longer time-scale,
spinning the star up to periods as short as a few ms (Alpar et
al.~1982). This model has recently gained strong support from the
detection of Doppler-shifted 2.49-ms X-ray pulsations from the
transient X-ray burster SAX J1808.4--3658 (Wijnands \& van der Klis
1998; Chakrabarty \& Morgan 1998).  At the end of the spin-up phase,
the secondary sheds its outer layers to become a white dwarf in orbit
around a rapidly spinning millisecond pulsar. A number of binary
millisecond pulsars now have compelling optical identifications of the
white dwarf companion.  The existence of solitary millisecond pulsars
in the Galactic disk (which comprise just under 20\% of all Galactic
millisecond pulsars) cannot easily be explained in the context of
this model and alternative formation scenarios need to be developed.

\section{Selection Effects}
\label{sec:selfx}

Having gotten a flavour for the likely evolutionary scenarios, we now
turn our attention to the first piece in the birth rate puzzle --- how
many active radio pulsars exist in the Galaxy? Although over 1200
sources are presently known, the total Galactic population is hidden
from us by a number of selection effects. As a result, the true number
of pulsars is likely to be substantially larger than the observed number.

\begin{figure}[hbt]
\setlength{\unitlength}{1in}
\begin{picture}(0,1.8)
\put(0.0,2.2){\includegraphics{xy.ps}}
\put(1.8,2.425){\includegraphics{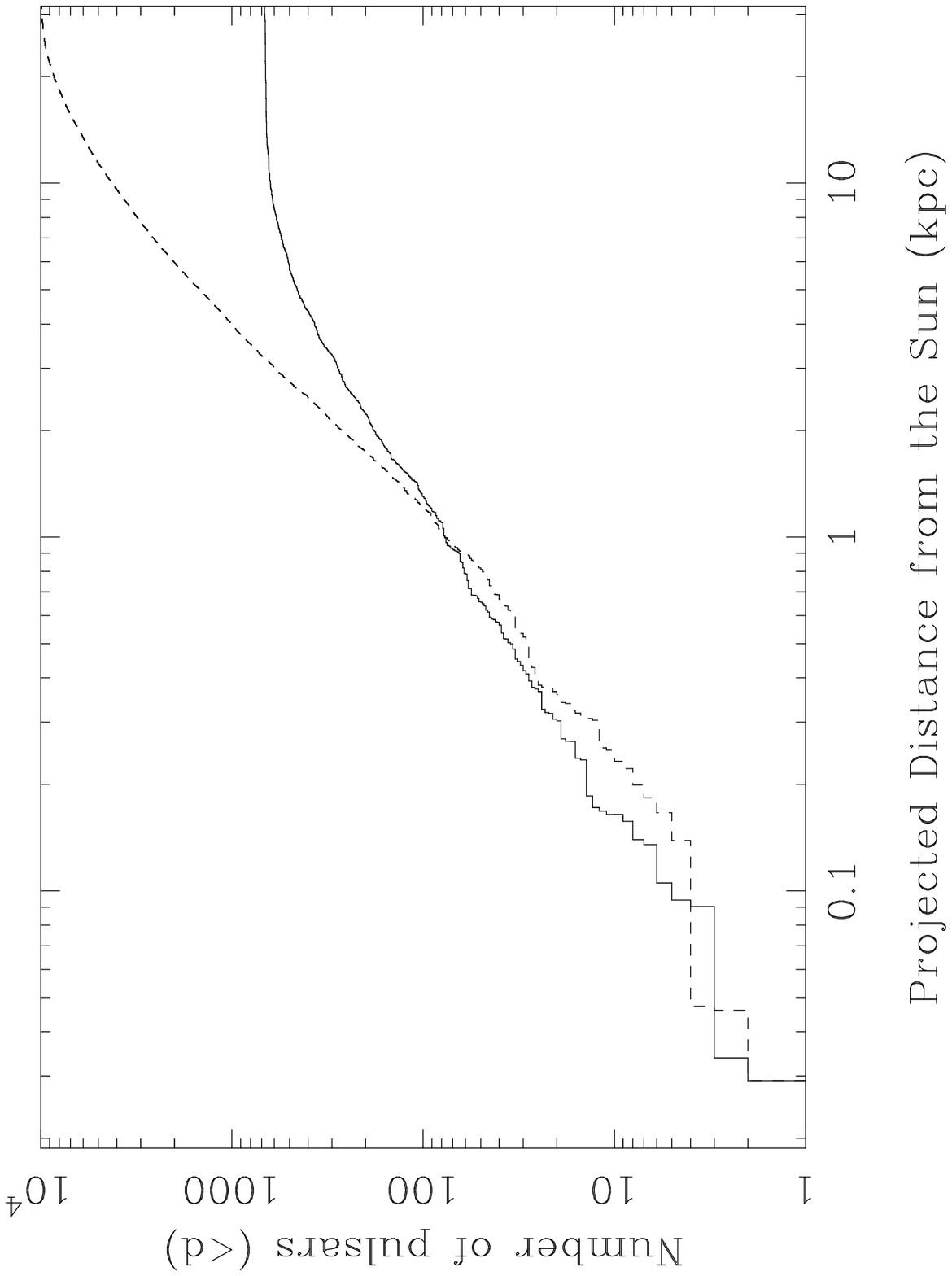}}
\end{picture}
\caption[]
{
Left: the observed sample of pulsars projected onto the Galactic
plane.  The Galactic centre is at (0,0) and the Sun is at (--8.5,0). 
Right: Cumulative number of observed pulsars (solid line) as a function 
of projected distance from the Sun.  The dashed line shows the expected 
distribution for a hypothetical uniform disk Galaxy (see text).
}
\label{fig:incomplete}
\end{figure}

The most prominent selection effect at play in the observed pulsar
sample is the inverse square law, i.e.~for a given luminosity the
observed flux density falls off as the inverse square of the
distance. This results in the observed sample being dominated by
nearby and/or bright objects. This effect is well demonstrated by the
clustering of known pulsars around our location when projected onto
the Galactic plane shown in Fig.~\ref{fig:incomplete}. Although this
would be consistent with Ptolemy's geocentric picture of the heavens,
it is clearly at variance with what we now know about the Galaxy, where
the massive stars show a radial distribution about the Galactic centre.  

The extent to which the sample is incomplete is shown in the
right panel of Fig.~\ref{fig:incomplete} where the cumulative number
of pulsars is plotted as a function of the projected distance from the
Sun. The observed distribution (solid line) is compared to the
expected distribution (dashed line) for a uniform disk population in
which there are errors in the distance scale, but no such selection
effects.  We see that the observed sample becomes strongly deficient
in terms of the number of sources for distances beyond a few kpc.

Beyond distances of a few kpc from the Sun, the apparent flux density
falls below the flux thresholds $S_{\rm min}$ of most surveys. Following 
Dewey et al.~(1984), we may parameterise the survey threshold by:
\begin{equation}
\label{equ:defsmin}
   S_{\rm min} = \frac{\beta \, \, \sigma_{\rm min} \,
   (T_{\rm rec}+T_{\rm sky})}
  {A_e \, \sqrt{N_p \, \Delta \nu \, t_{\rm int}}} \sqrt{\frac{W}{P-W}}.
\end{equation}
In this expression $\beta$ is a correction factor $\sim 1.3$ which
reflects losses to hardware limitations, $\sigma_{\rm min}$ is the
threshold signal-to-noise ratio (typically 7--10), $T_{\rm rec}$ and
$T_{\rm sky}$ are the receiver and sky noise temperatures (K), $A_e$
is the effective area of the antenna in K/Jy (1 K/Jy = 2760 m$^2$),
$N_p$ is the number of polarisations observed, $\Delta \nu$ is the
observing bandwidth, $t_{\rm int}$ is the integration time, $W$ is the
detected pulse width and $P$ is the pulse period.  This expression is
valid as long as $W < P$.

If the {\sl detected pulse width} $W$ equals the pulse period, the
pulsar is of course no longer detectable as a periodic radio source.
The detected pulse width is larger than the intrinsic value for a
number of reasons: finite sampling effects, pulse dispersion, as well
as scattering due to the presence of free electrons in the
interstellar medium. From Eq.~1 of my other article in this volume, it
is easy to show that the dispersive smearing scales as $\Delta
\nu/\nu^3$, where the bandwidth $\Delta \nu$ is assumed to be much
smaller than $\nu$, the observing frequency. As discussed in the
other article, this can largely be removed via the de-dispersion
process.  The smearing across the individual channels, however, still
remains and becomes significant when searching for short-period ($P
\lapp 200$ ms) pulsars located at large distances.

\begin{figure}[hbt]
\setlength{\unitlength}{1in}
\begin{picture}(0,2)
\put(0.9,-0.0){\includegraphics{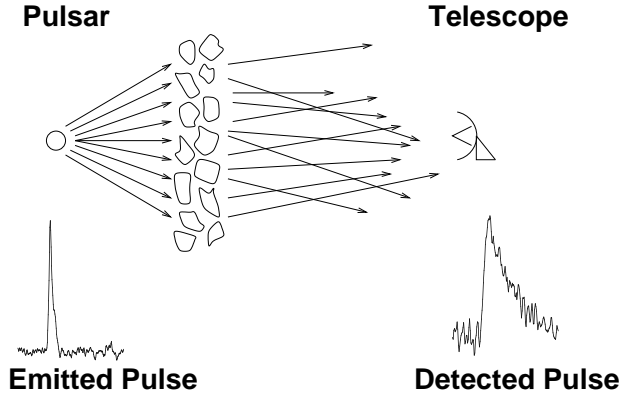}}
\end{picture}
\caption[]
{
Pulse scattering caused by irregularities in the interstellar medium.
The difference in path lengths and therefore in arrival times of the
scattered rays result in a ``scattering tail'' in the observed pulse
profile which lowers its signal-to-noise ratio.
}
\label{fig:scatt}
\end{figure}

As well as the dispersion broadening effect, free electrons in the
interstellar medium can scatter the pulses causing an additional
broadening due to the different arrival times of scattered pulses.  A
simple scattering model is shown in Fig. \ref{fig:scatt} in which the
scattering electrons are assumed to lie in a thin screen between the
pulsar and the observer (Scheuer 1968).
At observing frequencies $\lapp$ 400 MHz, scattering becomes
particularly important for pulsars with DMs $\gapp$ 200 cm$^{-3}$ pc.
The increased column density of free electrons can cause a
significant tail in the observed pulse profile as shown in
Fig.~\ref{fig:scatt}, reducing the effective signal-to-noise ratio and
overall search sensitivity. Multi-path scattering results in a
one-sided broadening due to the delay in arrival times which scales
roughly as $\nu^{-4}$. This cannot be removed by instrumental means.

Dispersion and scattering become most severe for distant pulsars in
the inner Galaxy as the number of free electrons along the line of
sight becomes large. The strong inverse frequency dependence of both
effects means that they are considerably less of a problem for surveys
at observing frequencies $\gapp 1400$ MHz compared to the usual 400
MHz search frequency.  An added bonus for such observations is the
reduction in $T_{\rm sky}$, since the spectral index of the
non-thermal emission is about --2.8 (Lawson et al.~1987).  Pulsars
themselves have steep radio spectra. Typical spectral indices are
--1.6 (Lorimer et al.~1995), so that flux densities are an order of
magnitude lower at 1400 MHz compared to 400 MHz. This can
usually be compensated somewhat by the use of larger receiver
bandwidths at higher radio frequencies. The tremendous success of the
multibeam surveys at Parkes is due to a combination of all these
factors which allow the survey to probe deeper into the Galaxy than
has been possible hitherto (Camilo et al.~2000).

A selection effect that we simply have to live with is beaming: the
fact that the emission beams of radio pulsars are narrow means that
only a fraction of $4\pi$ steradians is swept out by the radio beam
during one rotation.  A first-order estimate of the so-called
``beaming factor'' or ``beaming fraction'' ($f$) is 20\%; this assumes
a beam width of 10 degrees and a randomly distributed inclination
angle between the spin and magnetic axes.

\begin{figure}[hbt]
\setlength{\unitlength}{1in}
\begin{picture}(0,2.5)
\put(0.2,3.1){\includegraphics{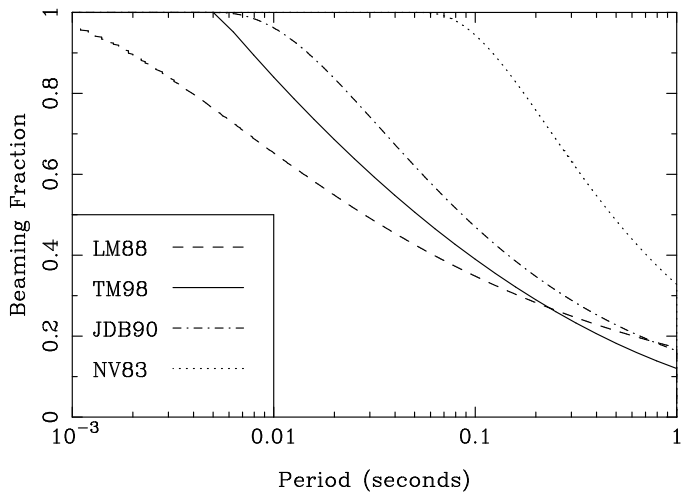}}
\end{picture}
\caption[]
{
Beaming fraction plotted against pulse period for four different
beaming models: Tauris \& Manchester (1998; TM88), Lyne \& Manchester
(1988; LM88), Biggs (1990; JDB90) and Narayan \& Vivekanand (1983; NV83).
}
\label{fig:bfracts}
\end{figure}

More detailed studies show that short-period pulsars have wider
beams and therefore larger beaming fractions than their long-period
counterparts (Narayan \& Vivekanand 1983; Lyne \& Manchester 1988;
Biggs 1990; Tauris \& Manchester 1998).  It must be said, however,
that a consensus on the beaming fraction-period relation has yet to be
reached.  This is shown in Fig.~\ref{fig:bfracts} where we compare
the period dependence of $f$ as given by a number of models.  Adopting
the Lyne \& Manchester model, pulsars with periods $\sim 0.1$ s beam
to about 30\% of the sky compared to the Narayan \& Vivekanand model
in which pulsars with periods below 0.1 s beam to the entire
sky. When many of these models were proposed, the sample of
millisecond pulsars was $\lapp$ 5 and hence their predictions about
the beaming fractions of short-period pulsars relied largely on
extrapolations from the normal pulsars.  A recent analysis of a large
sample of millisecond pulsar profiles (Kramer et al.~1998) suggests
that the beaming fraction of millisecond pulsars lies between 50 and 100\%.

Before concluding this section, it is appropriate to mention two other
selection effects which we will not discuss in detail: smearing of the
pulse due to motion in a binary system, and nulling of long-period
pulsars. The former effect is discussed in part in my other
contribution to these proceedings and is most relevant to the double
neutron star population (see Kalogera, this volume). The latter
effect, nulling, refers to the apparent spasmodic switch off in the
pulsar emission process first observed by Backer (1970) in a study of
single pulses at Arecibo.  The obvious connotation here is that
pulsars which spend significant periods in the ``null state'' will be
missed by any systematic searches, particularly those with short
integration times. For example, the existence of PSR B0826--34, was
extremely difficult to confirm since its null state often lasts many
hours (Durdin et al.~1979).

In the context of neutron star birth rates, the selection effect
caused by pulse nulling is of little significance, however, since it
is fairly certain from the work of e.g.~Ritchings (1976) that nulling
occurs predominantly in older, long-period pulsars.  In the context
of pulsar surveys in general, however, this effect is potentially
important. Nice (1999) has recently re-analysed archival Arecibo
search data to look for dispersed single pulses.  By comparison to the
Fourier-transform-based searches outlined in my other article, the
search analysis employed by Nice was trivial. One new 2-s pulsar,
J1918+08, that emitted only one detectable pulse (!) during the
original survey integration (67 s) was discovered. Nice's results
suggest that further long-period pulsars which only occasionally emit
detectable pulses may be found by future searches with only a modest
amount of additional effort.

\section{Accounting for Selection Effects}
\label{sec:corrfx}

Now that we know how severe the selection effects are, how do we
correct for them? There are three basic techniques: (1) source
counting; (2) Monte Carlo simulations; (3) scale factor calculations.

Counting sources (see e.g.~Kundt 1992) is by far the crudest method,
but nevertheless easy and instructive. In essence the trick is to look
at the cumulative distribution shown in Fig.~\ref{fig:incomplete} and
count sources out to a distance where you think the sample is
reasonably complete and then, by assuming some underlying distribution
function, extrapolate this number to get the total number of sources
in the Galaxy. Based on Fig.~\ref{fig:incomplete} we count 100 objects
out to 1 kpc, i.e.~a mean surface density of $100/(\pi \times 1 {\rm
kpc}^2) \simeq 30$ kpc$^{-2}$.  If pulsars have a radial distribution
similar to that of other stellar populations, the corresponding 
local-to-Galactic scale factor is then 1000$\pm$250 kpc$^2$ (Ratnatunga \&
van den Bergh 1989).  With this factor, we estimate there to be of
order 30,000 potentially observable pulsars in the Galaxy. Assuming a
beaming fraction of 20\% scales this to a total of 150,000.

Whilst this simple technique gives a rough answer, and can be done on
the back of an envelope, it is clearly not making good use of all the
available information contained in the observed sample. A considerably more
\begin{figure}[hbt]
\setlength{\unitlength}{1in}
\begin{picture}(0,3)
\put(1.1,-0.1){\includegraphics{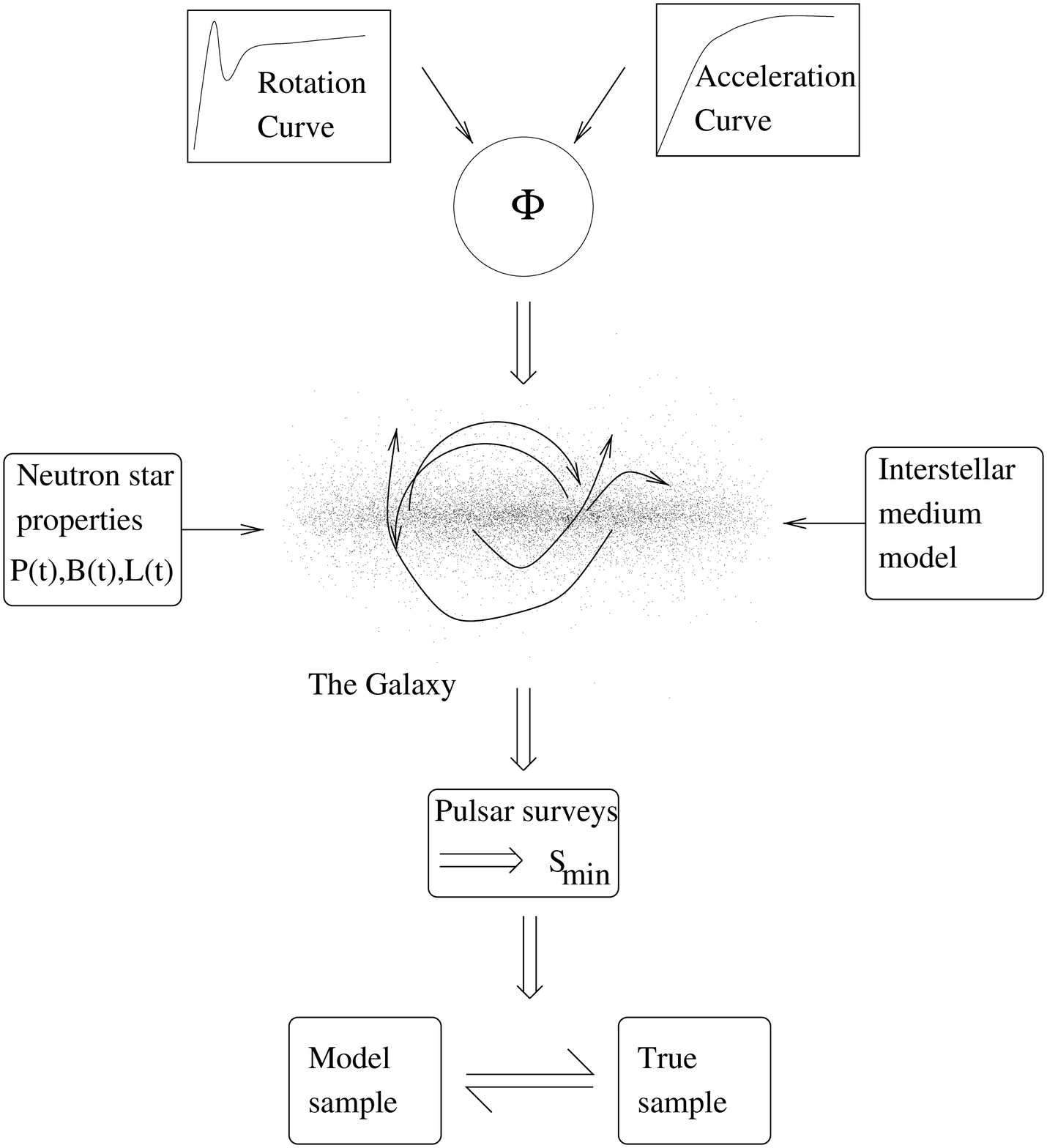}}
\end{picture}
\caption[]
{
The Full Monty --- an idealised neutron star population synthesis scheme.
}
\label{fig:fullmonte}
\end{figure}
rigorous approach, shown in Fig.~\ref{fig:fullmonte}, is to carry out
a Monte Carlo simulation of the Galaxy, the insterstellar medium and
the neutron star population.  Numerous accounts of this technique can
be found in the literature, see e.g.~Stollman (1987); Narayan (1987);
Emmering \& Chevalier (1989); Narayan \& Ostriker (1990); Bhattacharya
et al.~(1992).

In brief, the simulation seeds a model galaxy with pulsars having a
given set of initial parameters (period, magnetic field strength,
space velocity). The model pulsars are then allowed to evolve
kinematically in a model of the Galactic gravitational potential, and
rotationally (based on a model for pulsar spin-down and luminosity)
for a given time. The detailed models of the pulsar surveys then
produce a ``model observed sample'' which can then be directly
compared to the sample of 1200 objects that we actually observe.  By
varying the input model parameters and repeating the simulation to
maximise the agreement between the model and actually observed
samples, it is possible to directly constrain the physical parameters
of the Galactic neutron star population.

Although these Monte Carlo simulations assume a pulsar birth rate, and
in this sense can be used to obtain a ``best-fitting value'', problems
are often encountered in dealing with the large number of assumptions
required about pulsar spin-down and the radio luminosity evolution.
What these simulations do best is to teach us about selection effects
and allow useful tests of ideas hypothesised from the observable
population e.g.~magnetic field decay (Bhattacharya et al.~1992) and the
velocity-magnetic moment correlation (Lorimer, Bailes \& Harrison 1997).

The most model-independent way of constraining the number of pulsars
in the Galaxy and ultimately the birth rate is, following Phinney \&
Blandford (1981) and Vivekanand \& Narayan (1981), to define a scaling
function $\xi$ as the ratio of the total Galactic volume weighted by
pulsar density to the volume in which a pulsar could be detected by
various systematic surveys:
\begin{equation}
\xi(P,L) = \frac{\int \int_{\rm Galaxy} \, \Sigma(R,z) \, R \, \, dR \, dz}
           {\int \int_{P,L} \, \Sigma(R,z) \, R \, \, dR \, dz}.
\end{equation}
In this expression, $\Sigma(R,z)$ is the assumed pulsar density
in terms of galactocentric radius $R$ and height above the Galactic
plane $z$. Note that $\xi$ is primarily a function of period $P$ and
luminosity $L$ such that short period/low-luminosity pulsars have
smaller detectable volumes and therefore higher $\xi$ values than
their long-period/high-luminosity counterparts.

The scaling function is calculated in practice for each pulsar
individually using a Monte Carlo simulation to model the volume of the
Galaxy probed by the major surveys (Narayan 1987).  For a sample of
$N_{\rm obs}$ observed pulsars above a minimum luminosity $L_{\rm
min}$, the total number of pulsars in the Galaxy with luminosities
above this value is
\begin{equation}
N_{\rm gal} \approx \sum_{i=1}^{N_{\rm obs}} \frac{\xi_i}{f_i},
\end{equation}
where $f$ is the model-dependent beaming fraction discussed above.
The beaming fraction is, effectively, the only big unknown in this
calculation and it is often useful to quote potentially observable
pulsar number estimates (i.e.~those obtained before any beaming model
has been applied).

The most recent analysis to use the scaling function approach to
derive the characteristics of the true normal and millisecond pulsar
populations is based on the sample of pulsars within a cylinder of
radius 1.5 kpc centred on the Sun (Lyne et al.~1998).  The rationale
for this cut-off is that, within this region, the selection effects
are well understood and easier to quantify by comparison with the rest
of the Galaxy. These calculations should, at the very least, give
reliable estimates for the {\it local pulsar population}.

\begin{figure}[hbt]
\setlength{\unitlength}{1in}
\begin{picture}(0,1.6)
\put(-0.15,1.9){\includegraphics{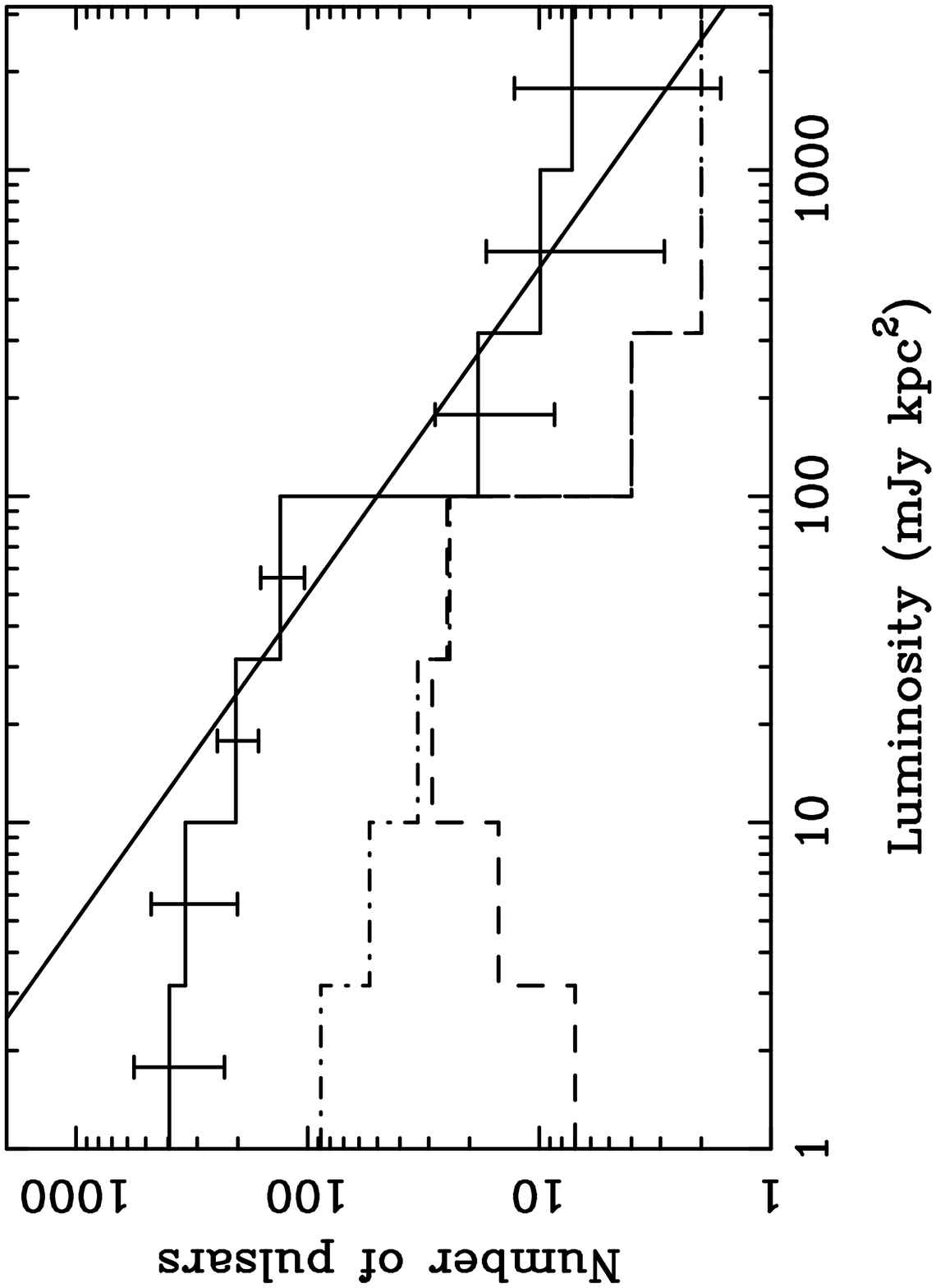}}
\put(+2.25,1.9){\includegraphics{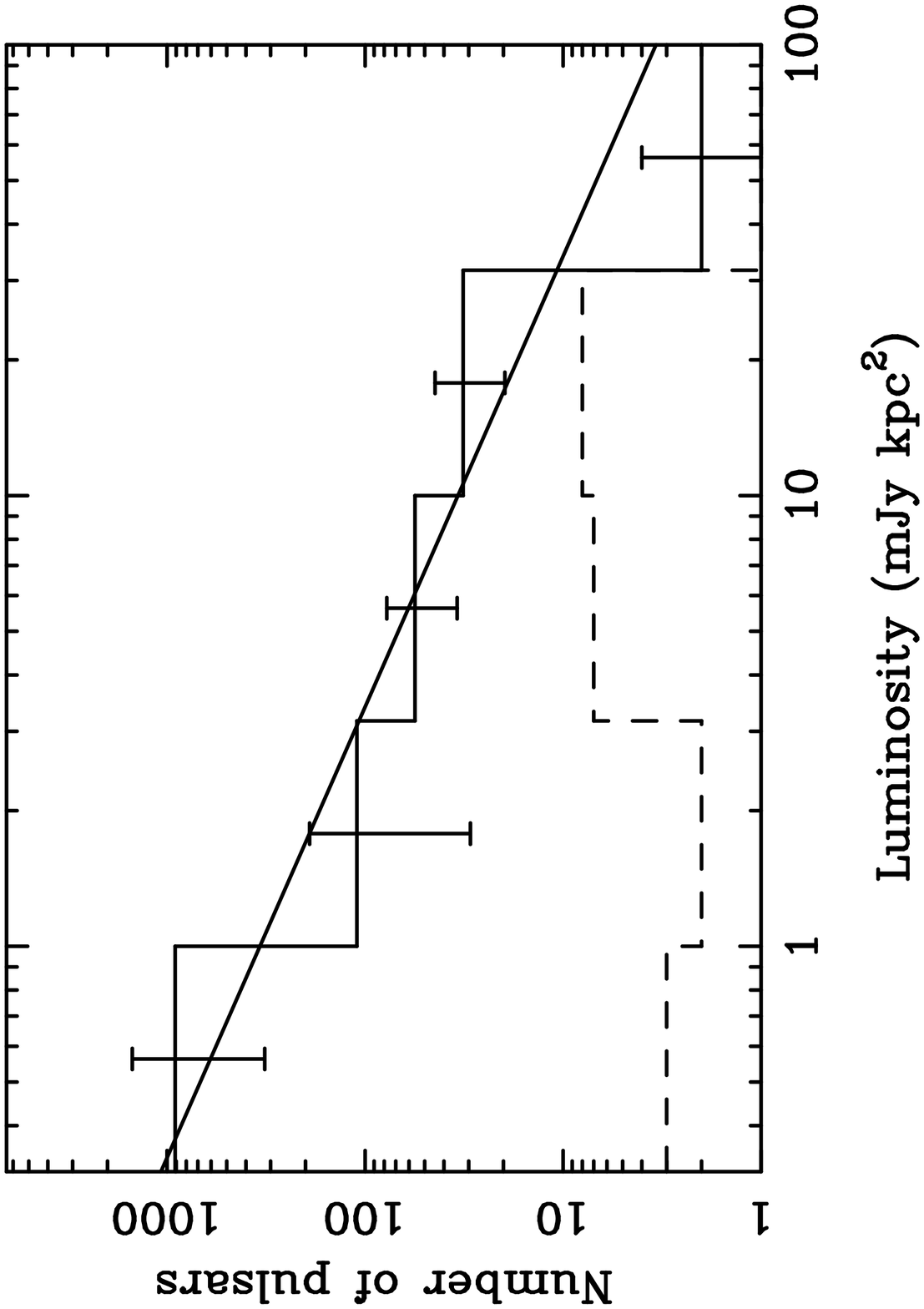}}
\end{picture}
\caption[]
{
Left: The corrected luminosity distribution (solid histogram with
error bars) for normal pulsars.  The corrected distribution {\sl
before} the beaming model has been applied is shown by the dot-dashed
line.  Right: The corresponding distribution for millisecond
pulsars. In both cases, the observed distribution is shown by the
dashed line and the thick solid line is a power law with a slope of
--1.  The difference between the observed and corrected distributions
highlights the severe under-sampling of low-luminosity pulsars.
}
\label{fig:lumfuns}
\end{figure}

The luminosity distributions obtained from this analysis are shown in
Fig.~\ref{fig:lumfuns}. For the normal pulsars, integrating the
corrected distribution above 1 mJy kpc$^2$ and dividing by $\pi
. (1.5)^2$ kpc$^2$ yields a local surface density, assuming Biggs'
(1990) beaming model, of $156 \pm 31$ pulsars kpc$^{-2}$. The
same analysis for the millisecond pulsars, assuming a mean beaming
fraction of 75\% (Kramer et al.~1998), leads to a local surface
density of $38 \pm 16$ pulsars kpc$^{-2}$ for luminosities above
1 mJy kpc$^2$.

Scaling these local surface densities of pulsars over the whole Galaxy
using the factor 1000$\pm$250 kpc$^2$ discussed above leads us to an
estimate of $(160\pm50)\times10^3$ active normal pulsars and
$(40\pm20)\times10^3$ millisecond pulsars in the Galaxy. These numbers
are to be compared with our back-of-the-envelope estimate of 150,000
pulsars (of all types) made at the beginning of this section. The robust
analysis using scale factors and more realistic beaming models
suggests a larger total population of objects.

\section{Birth Rates}
\label{sec:rates}

Simple birth rate estimates based on the above numbers and assuming
some mean lifetimes may now be made. For the millisecond pulsars, we
recall from the discussion in \S \ref{sec:psrpop} that their
characteristic ages are a few Gyr. This is corroborated by the
observations of white dwarf cooling ages and the effect of asymmetric
drift in the proper motions. Since millisecond pulsars cannot, by
definition, be older than the age of the Galactic disk itself (which
we take to be 10 Gyr; see e.g.~Jimenez, Flynn \& Kotoneva 1998 and
references therein) we set a lower limit on the millisecond pulsar
birth rate of $40,000/10^{10}=4\times10^{-6}$ yr$^{-1}$. We
discuss this in detail in \S \ref{sec:disc}.

For the normal pulsars, which are clearly much younger than the age of
the Galactic disk, we make use of a relatively model-free approach to estimate
the birth rate of normal pulsars that was first suggested by Phinney \&
Blandford (1981) and applied by Vivekanand \& Narayan (1981).  The
technique used the scale factor determinations described above to
calculate the flow of pulsars from short to long periods --- that is
the pulsar ``current''.

\begin{figure}[hbt]
\setlength{\unitlength}{1in}
\begin{picture}(0,1.2)
\put(0.2,-0.1){\includegraphics{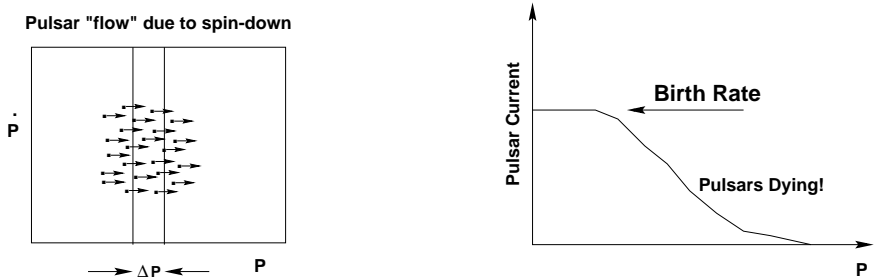}}
\end{picture}
\caption[]{
Schematic representation of pulsar current. Left: pulsar flow across the
$P-\dot{P}$ plane. Right: pulsar current (number of objects per unit time) 
as a function of period.
}
\label{fig:current}
\end{figure}

The relevance of pulsar current theory to birth rate calculations is
shown in Fig.~\ref{fig:current} --- a sketch of the expected situation
for a pulsar population born with short initial spin
periods. Following the initial ``plateau'', there is a steady decline
(reflecting luminosity and/or beaming evolution) until pulsars no
longer radiate sufficient energy to be detected. The birth rate of
this population is then simply proportional to the height of the
plateau. The major advantage of this approach over previous methods is
that it assumes nothing about the nature of pulsar spin--down and in
this sense is model-free. The only two assumptions underlying the
pulsar current theory are: (1) pulsars are spinning down to longer
periods; (2) the population has reached steady state i.e.~the mean age
of the population is much younger than the lifetime of the Galaxy.  It
should be clear by now that both these assumptions are justified for
normal pulsars in the disk of the Galaxy. As a result, the pulsar
current analysis does not require any {\it a-priori} knowledge of the
mean lifetime of the population to estimate the birth rate.

When Vivekanand \& Narayan (1981) first calculated the current, they
found a discontinuity in the distribution which showed a significant
increase in the current at a period of $\sim 0.5\,$s. They concluded
that, in order to account for this anomaly, a large number of pulsars
must be ``injected'' into the population with periods of about $0.5$
s, clearly challenging the conventional view that radio pulsars begin
their lives with short periods.  

A number of authors have investigated the injection issue further,
with contrasting results. Narayan (1987) and Narayan \& Ostriker
(1990) undertook detailed simulations and presented evidence in favour
of injection. Lorimer et al.~(1993), however, found little evidence to
support these claims.

Regardless of whether there is injection into the radio pulsar sample,
the discovery of long-period (5-12 s) anomalous X-ray pulsars in
supernova remnants (see e.g.~Mereghetti's review), which are
apparently radio-quiet, has important implications for the
spin periods and magnetic fields of neutron stars.  It is likely
that the anomalous X-ray pulsars will radically change our picture of
neutron star birth properties (Gotthelf \& Vasisht 2000)

\begin{figure}[hbt]
\setlength{\unitlength}{1in}
\begin{picture}(0,2.2)
\put(0.2,2.95){\includegraphics{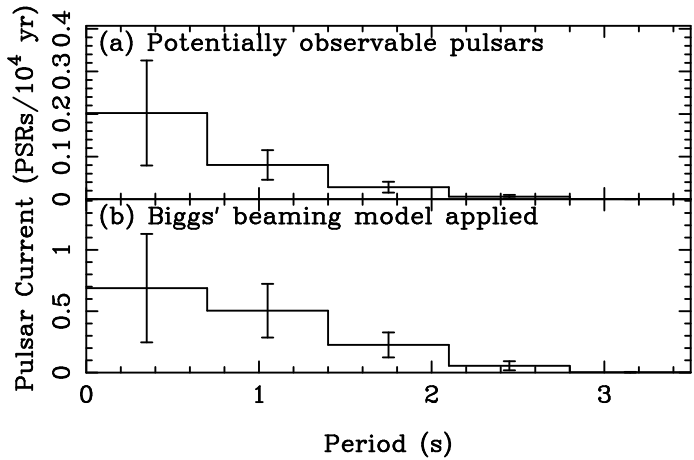}}
\end{picture}
\caption[]
{
The results of the pulsar current analysis carried out by Lyne et
al.~(1998). The top panel (a) shows the current derived with no
corrections for beaming. The lower panel (b) shows the data after
applying Bigg's (1990) beaming model.
}
\label{fig:lml+98}
\end{figure}

The results of the pulsar current analysis by Lyne et al.~(1998) are
shown in Fig.~\ref{fig:lml+98}. To derive the birth rate from this
figure, we take the pulsar current from the first bin (0.007 pulsars
per century assuming Biggs' beaming model) and divide this by $\pi
. (1.5)^2$ kpc$^2$ to scale it to unit surface area.  Multiplying this
number by the local-to-Galactic scale factor of $1000\pm250$ kpc$^2$
discussed above, the overall birth rate of normal pulsars is
$1.0\pm0.7$ pulsars per century, about 2500 times larger than the
lower limit obtained for the millisecond pulsars. This highly
significant difference in the birth rates of the two species is
further indication that the vast majority of binary systems disrupt at
the moment of the first supernova.

The relatively large error bars in the birth rate estimate for normal
pulsars reflects the model-free nature of the pulsar
current analysis adopted. Other authors (see e.g.~Narayan 1987) have
attempted to reduce the uncertainties by making additional assumptions
about pulsar luminosity. Whilst this is a commendable approach, there
is presently no convincing evidence to support these assumptions.  The
interested reader is referred to \S 4.2 of Lorimer et al.~(1993) for
further discussion which is not repeated here.

\section{Summary, Implications and Suggestions for Future Work}
\label{sec:disc}

Having made a lengthy tour, with many diversions and stops on the way,
we have finally achieved our goal of obtaining (some) birth rate
estimates for the neutron star population.  In this closing section,
we discuss the implications of these results in the context of the
proposed progenitor populations.

\subsection{Normal Pulsars and Core-Collapse Supernovae}

From the results of Lyne et al.~(1998), a normal pulsar is born on
average once every 60--330 yr in the Galaxy. More recently, Brazier \&
Johnston (1999) have derived a rate of one birth every 90 yr from a
local sample of X-ray detected neutron stars. This latter value is
independent of the radio pulsar beaming model. Based on an analysis of
historical supernovae by Tammann (see \S 8.3.2 of van den Bergh and
Tammann 1991) the rate of Galactic supernovae is one event every
10--30 yr. Thus, {\sl in the sense that the radio pulsar production
rate does not significantly exceed the supernova rate}, this is
consistent with our expectation from the standard model that every
normal radio pulsar we see began its life in a supernova.  Two
important caveats are intimately entwined with this statement:

\noindent
(1) The radio pulsar birth rate applies to objects with luminosities
above 1 mJy kpc$^2$, and needs to be boosted by some factor if a
significant part of the population is born with lower luminosities.
The study of Lyne et al.~(1998) shows that the radio luminosity
function is flattening between 1 and 10 mJy kpc$^2$
(Fig.~\ref{fig:lumfuns}).  Lorimer et al.~(1993) suggested that there
may be no need to have pulsars {\it born} with luminosities below 10
mJy kpc$^2$. The implication of this is that the faint pulsars have
undergone a significant luminosity evolution since birth. Further
studies of this issue are warranted.

\noindent
(2) The supernova rate is based on a sample which is a mixture of all
supernovae types. It should scaled by the fraction of core-collapse
supernovae. Based on the death rate of stars more massive than 6.5
M$_{\odot}$, From  the discussion in \S 8.5.1 of van den Bergh 
and Tammann (1991), one can deduce the Galactic rate of core-collapse
supernova to be event every 10--50 yr.

Assuming that few radio pulsars are born with exceedingly low
luminosities, it seems that the birth rate of radio pulsars is
significantly below the rate of core-collapse supernovae.  Presently,
there seems to be little evidence that the birth rates of soft
gamma-ray repeaters (Kouveliotou et al.~1994) and anomalous X-ray
pulsars (Gotthelf \& Vasisht 1998) will add significantly to the rates
discussed here, although further studies of the selection effects
imposed on these neutron star populations are warranted. Future
attempts to place tighter constraints on the pulsar birth rate (for
example based on a statistical analysis of the Parkes multibeam survey
results) should also clarify this situation.

\subsection{Millisecond Pulsars and low-mass X-ray binaries}

As derived in \S \ref{sec:rates}, based on the analysis of Lyne et
al.~(1998), and assuming a beaming fraction of 75\%, a lower limit to
Galactic birth rate of millisecond pulsars is $4\times10^{-6}$
yr$^{-1}$. This number is based on a sample of 18 sources within 1.5
kpc of the Sun with 400 MHz luminosities $\gapp 1$ mJy kpc$^2$.  The
entire sample from this paper is summarised in Table \ref{tab:lml+98}.
This includes the aforementioned objects, plus a further three
solitary millisecond pulsars with luminosities below 1 mJy kpc$^2$
that also lie within 1.5 kpc.

\begin{table}[hbt]
\begin{center}
\begin{tabular}{llllllll}
\hline
PSR & Type & $L_{400}$ & Scale & PSR & Type & $L_{400}$ & Scale \\
   &      & mJy kpc$^2$& Fact&      &     &mJy kpc$^2$& Fact \\
\hline
J0034$-$0534  & sB &   16  & 6  & J1744$-$1134  & Si & 0.4   & 210\\
J0437$-$4715  & sB &   19  & 3  & J1804$-$2717  & sB &   14  & 2  \\
J0711$-$6830  & Si &   11  & 7  & B1855$+$09    & sB &   12  & 3  \\
J1012$+$5307  & sB &   5   & 9  & J2019$+$2425  & lB &   2   & 40 \\
J1022$+$1001  & sB &   5   & 10 & J2033$+$1734  & lB &   10  & 6  \\
\\
J1024$-$0719  & Si & 0.6   & 180& J2051$-$0827  & sB &   9   & 7  \\
B1257$+$12    & Pl &   8   & 5  & J2124$-$3358  & Si & 0.4   & 520 \\
J1455$-$3330  & lB &   5   & 9  & J2145$-$0750  & sB &   18  & 2  \\
J1713$+$0747  & lB &   15  & 4  & J2229$+$2643  & lB &   13  & 5  \\
J1730$-$2304  & Si &   4   & 10 & J2317$+$1439  & sB &   68  & 2  \\
              &    &       &    & J2322$+$2057  & Si &   1   & 70 \\
\hline     
\end{tabular}
\end{center}    
\caption[]{
The sample of 21 millisecond pulsars used by Lyne et al.~(1998). For
each pulsar, we list the type of system, 400 MHz luminosity and
derived scale factor. Type is either: sB (short-period binary), lB
(long-period binary), Si (single pulsar) or Pl (planetary system).
The boundary between orbital periods of sB and lB systems is 25 days. Note that
this sample is confined to a cylinder of radius 1.5 kpc centred on the
Sun and does therefore not contain the few more distant millisecond
pulsars (e.g.~PSR B1937+21).
}
\label{tab:lml+98}
\end{table}

In the context of the standard formation scenario discussed in \S
\ref{sec:psrpop}, the binary millisecond pulsars descend from
low-mass X-ray binaries. The sum of the scale factors
of {\sl just} the binary systems (sB and lB types) listed in Table
\ref{tab:lml+98}, is 108. From the sample definition of Lyne et al.,
this translates to a local surface density of $108/(\pi.(1.5)^2)\simeq15$
binary millisecond pulsars kpc$^{-2}$.  Assuming, as before, that
these pulsars beam to 75\% of the sky, and using the Galactic scale
factor of 1000 kpc$^2$ (Ratnatunga \& van den Bergh 1989; see \S
\ref{sec:corrfx}), we estimate the Galactic population of binary
millisecond pulsars to be 20,000. Taking the maximum lifetime of this
population to be 10 Gyr results in a lower limit to the birth rate for
this population of $2\times10^{-6}$ yr$^{-1}$. This number is
comfortably below the birth rate for low-mass X-ray binaries ($\gapp 7
\times 10^{-6}$ yr$^{-1}$) derived by Cot\'e \& Pylyser (1989).

In the original millisecond pulsar birth rate analysis, Kulkarni \&
Narayan (1988) found a significant discrepancy between the birth rate
of short-orbital period ($\lapp 25$ day) millisecond pulsar binaries
and their progenitor low-mass X-ray binary systems, i.e.~the pulsar
birth rate was two orders of magnitude too large. Based on the larger
sample here, we can easily re-investigate this issue. Repeating the
above analysis for the 9 short-period (sB) binaries in Table
\ref{tab:lml+98}, we find the local surface density to be
$44/(\pi.(1.5)^2)\simeq6$ short-period binaries kpc$^2$.  Making the
same beaming and Galactic scaling correction as in the previous
calculation, we find a Galactic birth rate for these sources of $\gapp
8 \times 10^{-7}$ yr$^{-1}$. Whilst this is nominally a factor of at
least four larger than Cot\'e \& Pylyser's estimate for short-period
low-mass X-ray binary systems ($2 \times 10^{-7}$ yr$^{-1}$), the
discrepancy does not seem to be as pronounced as previously
thought. Based on the uncertainties in both birth rates (factors of a
few), one could conclude that the populations are consistent with each other.

Although all seems to be well with the connection between low-mass
X-ray binaries and millisecond pulsars, we should re-iterate the
statement made in \S \ref{sec:psrpop} that it is presently still
something of a mystery how the isolated millisecond pulsars are
formed. Based on the 6 single millisecond pulsars (Si) in Table
\ref{tab:lml+98}, we deduce a Galactic birth rate of $2\times10^{-5}$
yr$^{-1}$.  This is significantly larger than the birth rates for the
binary pulsars and is primarily a reflection of the low luminosities
observed for PSRs J1024--0719, J1744--1134 and J2124--3358 which
results in relatively large scale factor estimates.  Indeed, these
pulsars dominate the lowest bin in the luminosity function
(Fig.~\ref{fig:lumfuns}). It is presently marginally significant that
solitary millisecond pulsars have, on average, lower luminosities than
binary millisecond pulsars (Bailes et al.~1997; Kramer et
al.~1998). If confirmed by future discoveries, this may be a clue to
the origin of these mysterious objects.

As a final remark, we estimate from Table \ref{tab:lml+98} that the
Galactic population of millisecond pulsars with planetary systems is
of order 900. This estimate is based upon only one object and
therefore has a 100\% uncertainty!  It does appear, from the lack of
discoveries of similar systems, that the birth rate of pulsar
planetary systems in the Galactic disk is rather small.

\acknowledgements
I wish to thank the organisers of this meeting for putting together an
exciting programme, and providing a splendid venue in which to hold
it. Many thanks to Mike Davis and Fernando Camilo for extremely useful
comments on an earlier version of this manuscript.


\begin{thebibliography}{}

\bibitem[]{1}
Alpar~M.~A., Cheng~A.~F., Ruderman~M.~A., Shaham~J., 1982, Nat, 300, 728

\bibitem[]{2}
Backer~D.~C., 1970, Nat, 228, 42

\bibitem[]{3}
Bailes~M., 1989, ApJ, 342, 917

\bibitem[]{4}
Bailes~M. {\rm et~al.}, 1997, ApJ, 481, 386

\bibitem[]{5}
Bhattacharya~D., Wijers~R. A. M.~J., Hartman~J.~W., Verbunt~F., 1992, A\&A,
254, 198

\bibitem[]{6}
Biggs~J.~D., 1990, MNRAS, 245, 514

\bibitem[]{7}
Bisnovatyi-Kogan~G.~S., Komberg~B.~V., 1974, SA, 18, 217

\bibitem[]{8}
Blaauw~A., 1961, Bull. Astr. Inst. Netherlands, 15, 265

\bibitem[]{9}
Brazier~K. T.~S., Johnston~S., 1999, MNRAS, 305, 671

\bibitem[]{10}
Camilo~F. et al.~2000, Pulsar Astronomy --- 2000 and Beyond, (astro-ph/9911185)

\bibitem[]{11}
Chakrabarty~D., Morgan~E.~H., 1998, Nat, 394, 346

\bibitem[]{12}
Cot\'e~J., Pylyser~E. P.~H., 1989, A\&A, 218, 131

\bibitem[]{13}
Dewey~R. J.~et al.~1984, in
Reynolds~S., Stinebring~D., eds, Millisecond Pulsars.~p.~234

\bibitem[]{14}
Emmering~R.~T., Chevalier~R.~A., 1989, ApJ, 345, 931

\bibitem[]{15}
Flannery~B.~P., van~den Heuvel~E. P.~J., 1975, A\&A, 39, 61

\bibitem[]{16}
Gotthelf~E.~V., Vasisht~G., 1998, New Astronomy, 3, 293

\bibitem[]{17}
Gotthelf~E.~V., Vasisht~G., Pulsar Astronomy --- 2000 and Beyond, (astro-ph/9911344)

\bibitem[]{18}
Hansen~B., Phinney~E.~S., 1997, MNRAS, 291, 569

\bibitem[]{19}
Hills~J.~G., 1983, ApJ, 267, 322

\bibitem[]{20}
Hulse~R.~A., Taylor~J.~H., 1975, ApJ, 195, L51

\bibitem[]{21}
Jimenez~R., Flynn~C., Kotoneva~E., 1998, MNRAS, 299, 515

\bibitem[]{22}
Kouveliotou~C. et al.~1994, Nat, 368, 125

\bibitem[]{23}
Kramer~M. et al.~1998, ApJ, 501, 270

\bibitem[]{24}
Kulkarni~S.~R., Narayan~R., 1988, ApJ, 335, 755

\bibitem[]{25}
Kundt~W., 1992, in Hankins~T.~H., Rankin~J.~M., Gil~J., eds, 
{IAU} Colloquium 128.
Pedagogical University Press, Zielona G\'ora, Poland, p.~86

\bibitem[]{26}
Lawson~K.~D., Mayer~C.~J., Osborne~J.~L., Parkinson~M.~L., 1987, MNRAS, 225, 307

\bibitem[]{27}
Lorimer~D.~R., Bailes~M., Harrison~P.~A., 1997, MNRAS, 289,  592

\bibitem[]{28}
Lorimer~D.~R., Bailes~M., Dewey~R.~J., Harrison~P.~A., 1993, MNRAS, 263, 403

\bibitem[]{29}
Lorimer~D.~R., Yates~J.~A., Lyne~A.~G., Gould~D.~M., 1995, MNRAS, 273, 411

\bibitem[]{30}
Lyne~A.~G., Manchester~R.~N., 1988, MNRAS, 234, 477

\bibitem[]{31}
Lyne~A.~G. {\rm et~al.}, 1998, MNRAS, 295, 743

\bibitem[]{32}
Narayan~R., Ostriker~J.~P., 1990, ApJ, 352, 222

\bibitem[]{33}
Narayan~R., Vivekanand~M., 1983, A\&A, 122, 45

\bibitem[]{34}
Narayan~R., 1987, ApJ, 319, 162

\bibitem[]{35}
Nice~D.~J., 1999, ApJ, , 927

\bibitem[]{36}
Phinney~E.~S., Blandford~R.~D., 1981, MNRAS, 194, 137

\bibitem[]{37}
Ratnatunga~K.~U., van~den Bergh~S., 1989, ApJ, 343, 713

\bibitem[]{38}
Ritchings~R.~T., 1976, MNRAS, 176, 249

\bibitem[]{39}
Scheuer~P. A.~G., 1968, Nat, 218, 920

\bibitem[]{40}
Shapiro~S.~L., Teukolsky~S.~A., 1983, The physics of Compact Objects. 
Wiley, New York

\bibitem[]{41}
Smarr~L.~L., Blandford~R., 1976, ApJ, 207, 574

\bibitem[]{42}
Spruit~H., Phinney~E.~S., 1998, Nat, 393, 139

\bibitem[]{43}
Stollman~G.~M., 1987, A\&A, 178, 143

\bibitem[]{44}
Tauris~T.~M., Manchester~R.~N., 1998, MNRAS, 298, 625

\bibitem[]{45}
Toscano~M., Sandhu~J.~S., Bailes~M., Manchester~R.~N., Britton~M.~C.,
Kulkarni~S.~R., Anderson~S.~B., Stappers~B.~W., 1999, MNRAS, 307, 925

\bibitem[]{46}
van~den Bergh~S., Tammann~G.~A., 1991, ARAA, 29, 363

\bibitem[]{47}
van Kerkwijk~M.~H., 1996, in Johnston~S., Walker~M.~A., Bailes~M., eds,
  Pulsars: Problems and Progress, {IAU} Coll 160.
\newblock Astronomical Society of the Pacific, San Francisco, p.~489

\bibitem[]{48}
Vivekanand~M., Narayan~R., 1981, JAPAS, 2, 315

\bibitem[]{49}
Wijnands~R., van~der Klis~M., 1998, Nat, 394, 344

\bibitem[]{50}
{Young}~M.~D., {Manchester}~R.~N., {Johnston}~S., 1999, Nat, 400, 848

\end{thebibliography}
\end{document}